# Discovery of Soliton Self-Mode Conversion in Multimode Optical Fibers

L. Rishøj[1], B. Tai[1], P. Kristensen[2], S. Ramachandran[1,*]

[1]*Boston University, 8 St Mary's St, Boston, MA 02215, USA.*

[2]*OFS-Fitel, Priorparken 680, Brøndby 2605, Denmark.*

*\*Corresponding author:* [sidr@bu.edu](mailto:sidr@bu.edu)

**Abstract**

Nonlinear optical wave propagation manifests in a multitude of frequencies generated from quantum-noise, and selecting desired nonlinear products usually requires seeding the medium with extraneous waves, employing spatial or spectral filters, or operation in resonant cavities. This is especially true for multimode systems because of their high density of states. Here we report the discovery of a self-action effect, originating from quantum noise, leading to complete nonlinear optical conversion of an ultrashort soliton pulse between two distinct, spatially coherent eigenmodes that are frequency-separated by one Raman Stokes shift. That systematic nonlinear spatial reconfiguration occurs in fibers with mode counts exceeding 10,000, which are often deemed to be chaotic, points to the fundamental role of *intermodal* group velocity dispersion in the selection rules for multimode nonlinear optics. We demonstrate wideband spectral translations of ~70% of the carrier frequency, and the generation of record, Megawatt peak-power pulses in the biologically crucial 1300-nm spectral window, directly out of a flexible optical fiber. More generally, this novel nonlinear coupling mechanism may be applied to fibers or on-chip waveguides, and facilitate, power-scalable spatially coherent, ultrashort pulse generation from the visible to the mid-IR.

## 1. Introduction

The nonlinear dynamics of light propagation through a medium critically depends on the group velocity and its spectral gradient, group velocity dispersion (GVD), experienced by the pulse. The existence of a medium with anomalous GVD is a necessary condition for obtaining solitary pulse propagation. The propagation speeds of solitons may also be influenced by interactions with other solitons, because of attractive potentials for one pulse created by cross-phase modulation (XPM) from the other soliton (*1*,*2*,*3*,*4*). Alternatively, the nonlinearity of a medium can be used to control the group velocity of a pulse, as with self-phase matched nonlinearities such as stimulated Brillouin scattering or stimulated Raman scattering (SRS) to achieve slow or fast light (*5*,*6*). While the centrality of group velocity in tailoring the nonlinear dynamics of pulses is well established for bulk media and single mode fibers (SMF) or waveguides, little attention has been paid to the opportunities as well as challenges in exploiting or tailoring these parameters in multimode systems.

Nonlinear optics in multimode optical fibers, although studied since the 1970s (*7*), has been the subject of resurgent interest because linear light propagation in multimode fibers has recently been shown to be of immense utility in areas such as imaging (*8*), information capacity scaling of optical networks (*9*), and high-power fiber laser development (*10,11*). Broadly, work in this area can be divided into two categories: The enhanced density of states available in a multimode fiber enables greater flexibility in phase matching, implying that four-wave mixing (FWM) interactions are possible over significantly larger spectral ranges than that typically afforded by SMF (*7,12,13*). Alternatively, since GVD depends on mode order, modal choice along with waveguide design provides a larger design space to tailor nonlinear coupling (*14,15,16,17,18*). The second category involves graded index (GRIN) fibers that were theoretically postulated in the 1980s to yield multimode soliton entities (*19,20*). Recent experimental work has validated these postulates (*21*) and revealed interesting spatio-temporal effects (*22,23*). In GRIN fibers, the group velocities of all modes are nearly identical, and as such they behave substantially like bulk media. Likewise, reported observations of nonlinear optical effects include phenomena similar to those in bulk media or SMF, such as cascaded stimulated Raman scattering (*24*) and soliton self-frequency shifting (SSFS) (*22*).

Here, we study the influence of a multimode fiber's intermodal group velocity on nonlinear scattering, and discover that, when certain conditions for group velocity distributions are satisfied, a unique nonlinear coupling phenomenon manifests, with no direct analogue in bulk or guided wave systems, to the best of our knowledge. Given the large density of states in a multimode fiber, two distinct spatial modes*, and only those two modes*, may have identical group velocities at specific spectral separations. When this spectral separation corresponds to the frequency difference at which nonlinear gain is maximized, efficient coupling between the modes may be expected. Specifically, we discover that when this condition is satisfied for Raman gain, a spectrally separated new pulse in a different mode initiates from quantum noise, but owing to the strong coupling *and* group-velocity matching, results in the transfer of 99% of the photons from the original pulse to this new pulse (*25*). Raman gain mediated by group velocity matching between modes has been observed birefringent fibers and ring resonators before, but only when both interacting pulses were seeded (*26*) or when a resonant cavity was employed (*27*). We observe this effect, called soliton-self mode conversion (SSMC) henceforth, in step index fibers with mode counts exceeding 10,000. However, the system self-selects precisely one nonlinear process out of the multitude of possibilities, without any seeding or feedback with resonant cavities or filters. Efficient and exclusive coupling of this sort requires symbiotic feedback between two spectrally separated soliton pulses, which means that group velocities need to be matched, *and* the anomalous GVD condition needs to satisfied at both wavelengths. This is impossible in a bulk medium, not known to be possible to engineer in SMFs even with the advent of photonic crystal designs, and is thus unique to multimode fibers. More generally, this points to the powerful role group-velocity distributions play in multimode nonlinear fiber optics, and suggests that engineering this parameter would lead to the observation of new nonlinear coupling effects, such as the SSMC phenomenon reported here.

## 2. Observation of SSMC

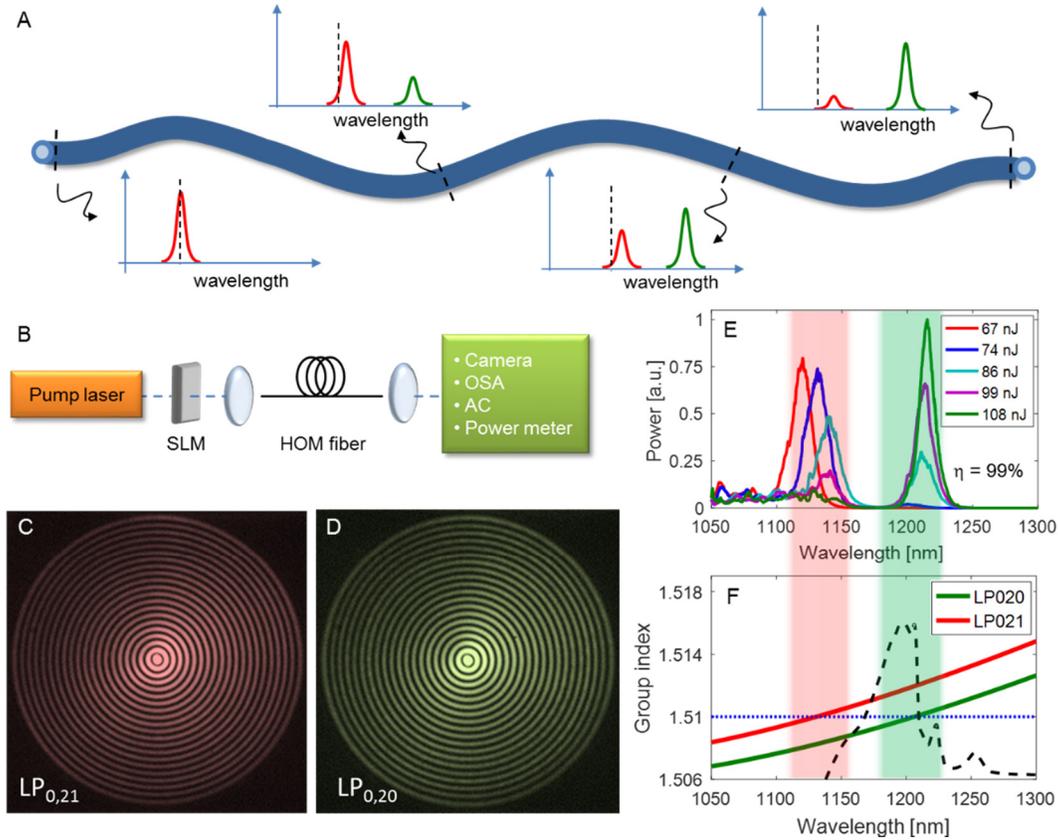

**Fig. 1 Demonstration of soliton self-mode conversion.** (**A**) Schematic illustration of pulse evolution along a fiber via soliton self-mode conversion (SSMC). The red and green colored pulses denote different spatial modes and the dashed black line marks the center wavelength of the input pulse. As the input soliton pulse (red) propagates along the fiber, it generates a new pulse (green) via Raman scattering in a different spatial mode that is group-index matched to the original pulse. Further propagation completely depletes the original pulse, resulting in transfer of power from one mode to another via this self-organized nonlinearity. (**B**) Schematic of the experimental setup. (**C**) Experimental image of the $LP_{0,21}$ mode and (**D**) $LP_{0,20}$ mode. (**E**) Experimental demonstration of the SSMC process. Energy transfer occurs between two group index matched modes, the spectral components in the red and green shaded boxes are in the $LP_{0,21}$ and $LP_{0,20}$ mode, respectively. (**F**) Group index as a function of wavelength for the two modes, black dashed line indicates the Raman gain coefficient. This illustrates that SSMC is the preferred nonlinear phenomenon when group-index matched spectral separation matches with the Stokes shift of Raman scattering.

Consider a soliton pulse, with ~17 nm spectral bandwidth at ~1120 nm, in the $LP_{0,21}$ mode (measured mode image in Fig. 1C) at the input of a step index silica MMF with 0.34 numerical aperture (NA) and 97 μm core diameter (the schematic of the experimental setup is shown in Fig. 1B, the desired mode is excited in the higher order mode (HOM) fiber by encoding the spatial phase onto a Gaussian beam using a spatial light modulator (SLM); more details about the experimental setup, fiber characteristics, and the mode excitation method is provided in the supplementary material). This fiber supports ~10,735 spatial eigenmodes, but since the $LP_{0,m}$ modes are increasingly more linearly stable to bend perturbations as the radial order *m* increases (28) (see supplementary materials Fig. S1B), and they also possess anomalous GVD (*11*), a pulse in this mode is expected to behave like any soliton and experience frequency shifts due to SSFS as input power or fiber length increases (*29,30*). Here, a rather unexpected phenomenon occurs – Fig. 1E shows the spectra at the output of the fiber measured for increasingly higher launched pump pulse energies. The soliton does indeed appear to shift its frequency by the conventional SSFS effect, but starts losing energy to a new spectral feature, spontaneously formed at a spectral separation of ~15 THz. For sufficient input power, complete conversion of the photons in the original pulse to this new pulse occurs. Imaging the output of the fiber with appropriate spectral filters reveals that this spectral feature is in a different spatial eigenmode, the $LP_{0,20}$ mode (Fig. 1D), which is exactly one radial order lower than the input mode. We systematically imaged all the spectral features in the red and green wavelength windows, and confirmed that they correspond to the $LP_{0,21}$ and $LP_{0,20}$ modes, respectively. Figure 1F shows that the group index at 1140 nm of the input $LP_{0,21}$ mode is identical to that of the $LP_{0,20}$ mode at 1210 nm, and this frequency separation is close to the peak of the Raman gain coefficient – i.e. one Raman Stokes shift away. Thus, SSMC represents the complete conversion of an ultrashort pulse in one mode into an ultrashort pulse in another mode, when the spectral separation of two group-velocity-matched modes is close to maximum Raman gain. Note that this process (schematically illustrated as a function of fiber length in Fig. 1A) wins over conventional SSFS, although the latter is a self-seeded process that is typically expected to dominate over noise-initiated nonlinear coupling phenomena.

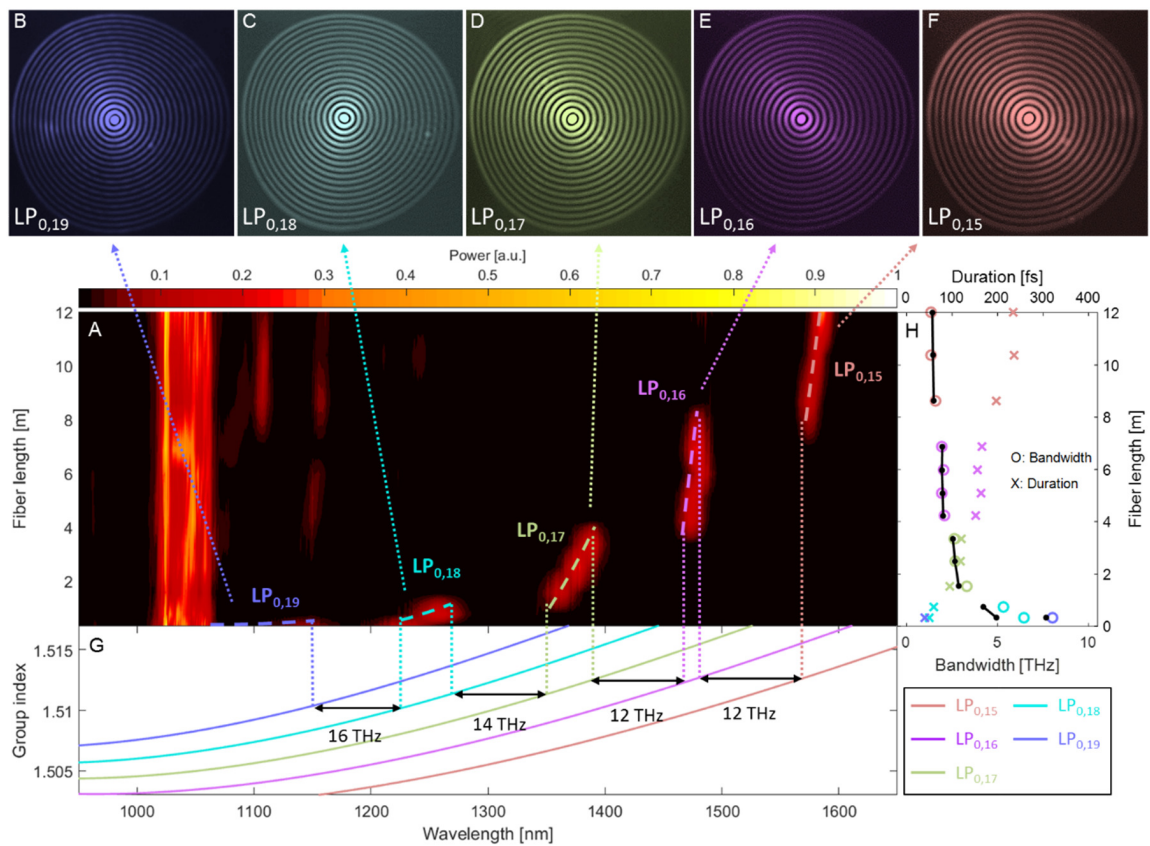

**Fig. 2 Evolution of SSMC along fiber.** (**A**) Spectra as a function of fiber length, the launched pulse at 1045 nm is in the $LP_{0,19}$ mode. (**B**)-(**F**) Representative mode images taken using 10 nm bandpass filters of various spectral features. (**G**) Simulated group indices for the modes of interest, frequency separation for various SSMC processes indicated on plot. (**H**) Measured spectral bandwidth and calculated transform limited temporal duration as a function of fiber length of different pulses. Colors denote various modes and the black line is the theoretical bandwidth.

This competition between SSMC and conventional SSFS is captured well by measuring the full pulse evolution in a similar step-index fiber (NA~0.34; core-diameter ~87 µm). For this experiment a 100-fs pulse with 72.4-nJ energy at 1045 nm (KMLabs, Y-Fi) is used to excite a pure $LP_{0,19}$ mode at the fiber input. Figure 2A shows the spectral evolution with propagation length obtained via cutback of the original 12 m long fiber, along with measured mode images (Fig. 2B-F). Figure 2G shows the simulated group index for the participating modes and the associated frequency shifts for each SSMC process. Initially, a soliton formed in the $LP_{0,19}$ mode red-shifts via conventional SSFS to 1141 nm after 33 cm of propagation. At this point, it is group index matched to the $LP_{0,18}$ mode at 16 THz frequency separation (i.e. within the Raman gain bandwidth). Hence, power transfer from the $LP_{0,19}$ pulse to the vastly wavelength separated $LP_{0,18}$ mode commences, and after ~40 cm of propagation, the mode and wavelength conversion is complete. Thereafter, the new fundamental soliton in the $LP_{0,18}$ mode shifts to 1257 nm via conventional SSFS, when it becomes group index matched to the $LP_{0,17}$ mode at 14 THz frequency separation. As in the previous case, with further propagation, full power transfer to the $LP_{0,17}$ mode is achieved via SSMC. These alternating intra-modal (SSFS) and intermodal (SSMC) processes repeat along the fiber – at the longest tested fiber length, a spatially coherent pulse at 1587 nm in the $LP_{0,15}$ mode is obtained. A similar effect is expected for power tuning, as illustrated for one specific mode pair in Fig. 1. Indeed, with higher input power, we observed the formation of the next ($LP_{0,14}$) mode, at an even longer wavelength (1696 nm), but we were constrained from increasing pump power further to obtain full photon conversion (spectrum in Fig. S5 of supplementary materials). Figure 2H shows the measured spectral bandwidths and corresponding transform-limited pulse durations versus fiber length. The black lines represent the theoretical bandwidth based on the assumption that all pulses are $1^{st}$ order solitons, and that the photon number is conserved during the fiber cutback. This bandwidth is given by

$$\Delta v \propto \frac{1}{D \cdot A_{eff} \cdot \lambda^4}, \qquad (1)$$

where $D$ and $A_{eff}$ are the group velocity dispersion and effective area for the various modes. Since the experimental data fall on this line, we conclude that all SSMC processes indeed conserves photon number after the initial $LP_{0,19}$ soliton is formed from the pump pulse.

## 3. Pulse Characterization

In order to characterize the dynamics of the SSMC process further, a second set of experiments were performed with a 370-fs pump laser exciting a $LP_{0,21}$ mode in a 97 µm core diameter fiber (NA ~ 0.34). The output spectra for increasing pump power are presented in Fig. 3. In the first three traces (blue, green, and red) a fundamental soliton is formed and red-shifts due to conventional SSFS. In the red trace a secondary 1st order soliton appears at 1072 nm, due to conventional soliton fission because of the longer pulse duration of our pump laser. Once the fundamental soliton reaches ~1184 nm (cyan trace) it has a pulse energy of 64 nJ, and based on its autocorrelation measured pulse width of 59 fs, the estimated peak power is 1.1 MW. The onset of SSMC is evident from the appearance of a weak spectral feature at 1260 nm (in the bottom of the green shaded box). Further pump power increase transfers power from the pulse tracked by the red shaded box to the SSMC feature in the green shaded box. The shifted soliton is in the $LP_{0,21}$ mode – same as the pump mode, whereas the newly created spectral feature (green shaded region) is in the $LP_{0,20}$ mode. In the, highest power, black trace, most of the power from the soliton is transferred to the SSMC generated pulse at 1317 nm in the $LP_{0,20}$ mode. This pulse was spectrally isolated using a longpass filter, and autocorrelation measurements revealed a pulse width of 74 fs, which is close to the transform limited pulse width of 65 fs calculated from its spectral bandwidth. As the measured pulse energy is 80 nJ this represents a peak power of 1.1 MW, directly out of the fiber. Note the additional spectral feature at 1224 nm in the black trace. This is also in the $LP_{0,20}$ mode, but has no relevance to the SSMC process described above, since it arises from a SSMC process involving the secondary soliton described earlier.

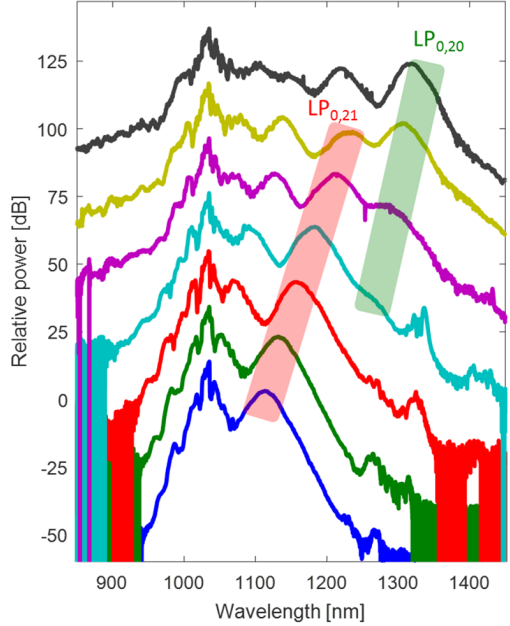

**Fig. 3 Power tuning of SSMC.** Spectra for increasing pump power (offset for clarity), the red and green shaded boxes indicate the spectral regions where the spatial mode is confirmed to be the $LP_{0,21}$ and $LP_{0,20}$ modes, respectively. The fiber length is 54 cm.

Figure 4A is a plot of pulse energy pump power. For relative pump powers < 4 dB, conventional SSFS yields a pulse-energy-vs-pump-power slope of 5.6 nJ/dB. Between relative pump power levels ~4.5-5.5 dB the SSMC process occurs with a dramatically different slope of ~92.3 nJ/dB. Once full conversion transpires, the SSMC generated pulse undergoes traditional SSFS again, as evident from an energy-vs-pump-power slope (6.2 nJ/dB) similar to that of the original soliton. Autocorrelation traces of the lower wavelength pulse in the $LP_{0,21}$ mode prior to the initiation of SSMC reveal pulse durations of ~60 fs, whereas the longer wavelength pulse in the $LP_{0,20}$ mode during and after SSMC are ~74-78 fs wide (31). The different pulse widths of the soliton before and after SSMC are consistent with the fact that the fiber parameters (GVD of the mode $D$, mode area $A_{eff}$ – see supplementary materials Fig. S3) are different for the different modes and wavelengths, and the need for a photon-conserving process to maintain soliton number $N$, given by

$$N^2 = \frac{4\pi^2 c n_2 P T_0^2}{D \lambda^3 A_{eff}}, \qquad (2)$$

where $n_2$ is the nonlinear coefficient, $P$ is the power, and $T_0$ is the pulse duration. Figure 4B shows that the normalized soliton number for the SSFS soliton is nominally constant, as expected. However, the SSMC generated pulse, while always transform-limited regardless of its energy, initially has a normalized soliton number of only 0.3 before rapidly approaching unity. This indicates that the SSMC pulse is not a traditional soliton, and instead a combination of intermodal cross-phase modulation and Raman gain serves to preserve its pulse width. Note that, whenever the conditions for SSMC are met, either for the single transition depicted in Fig. 3, or the multiple SSMC transitions shown in Fig. 2, full conversion occurs and photon number is conserved. Hence SSMC is the preferred nonlinearity over the self-seeded, conventional SSFS process. Since SSFS frequency shifts scale with pulse width as $1/T_0^4$, SSFS may be expected to subsume SSMC for extremely short pulses. Nevertheless, our observations indicate that SSMC is robustly the preferred nonlinearity even for pulses as short as ~60 fs.

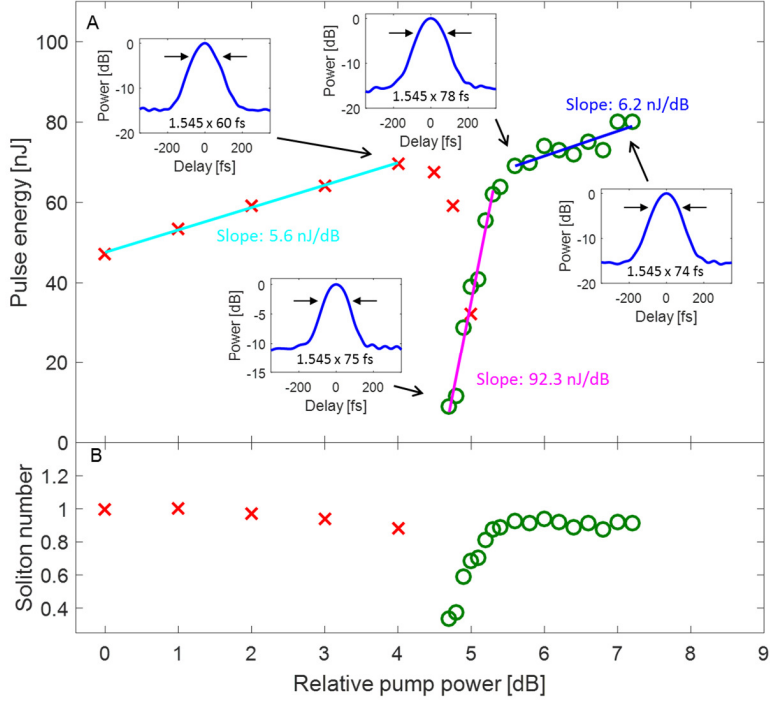

**Fig. 4 Dynamics of SSMC.** (**A**) Pulse energies for the pulse in the $LP_{0,21}$ mode (red crosses) and the SSMC pulse in the $LP_{0,20}$ mode (green circles) as a function of relative pump power. Linear fits to the various slopes are provided on the plot. The inserts are autocorrelation data for four selected pulses. (**B**) Soliton number normalized with respect to the pulse at the lowest pump power.

The mode image of the 1.1 MW peak power pulse at 1317 nm obtained in the black trace of Fig. 3 (and final data point in Fig. 4) is shown in Fig. 5B. The seemingly pure $LP_{0,20}$ mode image suggests that the output is a high quality spatially coherent and *single-moded* beam. We quantify the mode quality via axicon-lens based conversion of the output beam into a Gaussian shaped beam, using the experimental setup shown in Fig. 5A. This schematic should yield a perfect Gaussian with a theoretical transmission of 86% if the output was indeed single-moded (simulations are provided in supplementary in Fig. S4) (*32*). The fact that we obtain a high quality Gaussian beam (image in Fig. 5C, fit in Fig. 5D) with a measured transmission of 81% – close to theoretical prediction – proves that SSMC yields pure and spatially coherent modes. Thus, the SSMC output may be used as is, since it resembles a truncated Bessel beam (*33*) that has found several beneficial applications in microscopy (*34*), or may be converted into a conventional Gaussian beam with minimal loss.

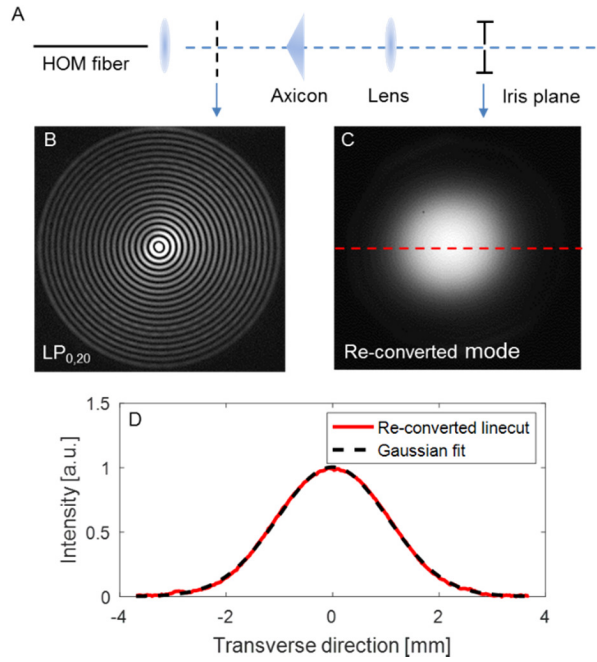

**Fig. 5 Spatial mode purity characterization.** (**A**) Experimental setup for mode re-conversion. (**B**) Image of the $LP_{0,20}$ output mode. (**C**) Gaussian-like beam re-converted using axicon after spatial filtering. (**D**) Linecut of re-converted beam and corresponding Gaussian fit proving the high spatial coherence of modes obtained via SSMC.

## 4. Discussion

The soliton self-mode conversion (SSMC) effect we have described here is a nonlinear scattering mechanism that fully converts an ultrashort soliton in one spatial mode into a group-velocity-matched soliton in a distinct spatial mode. The driving mechanism is the existence of substantial Raman gain for a noise initiated pulse that, despite being at large spectral separation from the "pump" pulse, has identical group velocity as well as anomalous GVD. The interesting observation is that a system comprising more than 10,000 spatial modes self-selects precisely one nonlinear process, initiated from quantum noise, and facilitates full photon conversion. In doing so, it even subsumes the competing nonlinearity of self-seeded Raman soliton shifting (SSFS) that is prevalent for intense ultrashort pulses in anomalous GVD fiber. In contrast, most other nonlinear coupling phenomena become dominant and offer full pump depletion or complete conversion only when the process is externally seeded, as with parametric converters (*35*), self-seeded, as with SSFS (*36*) and self-focusing (*37*), constrained by filtering, either spectrally as with similariton fiber lasers (*38,39*) or temporally as with passive mode-locking using a saturable absorber (*40*), or operated in a resonant cavity, as with parametric oscillators (*41*). Even so, the nonlinearly generated wave often acts as a secondary pump for cascaded nonlinear effects, though that in itself can be interesting, such as in generating supercontinua (*42*) or multicasting data (*43*). Thus, SSMC appears to be a process that has no direct analogues to nonlinear effects commonly observed in bulk media or SMF, and is unique to multimode waveguides, provided that group-index spacings between modes supply the selectivity.

We used step-index multimode fibers for our experiments because they possess several attributes that are attractive for this and many other nonlinear optical studies. It is well known that propagation constant, and hence group-velocity, spacings are irregular in step-index fibers, which means that spectral separations at which the group velocities are matched between modes changes with mode order. In addition, light propagation stability in step-index fibers increases with radial-order for zero-angular-momentum modes (the so-called $LP_{o,m}$; m ∈ radial order) (*28*). Finally, modes in step-index fibers possess anomalous GVD (*11*) and large mode areas (*10*). Hence, step index fibers yield a power scalable platform in which GVD *and* intermodal group-velocity can be tailored by mode choice and fiber core size. Using SSMC, we demonstrate ~700 nm wavelength translations in the 1000-2000-nm spectral range, as well as the ability to obtain record (> 1 MW) peak powers at the biologically crucial ~1300 nm wavelength range (*44*), directly out of flexible fibers for the first time. Note that this marks the first source that facilitates behavioral deep neuroimaging of freely moving animals using tethered microscopes, since the fiber is long and flexible, and no (grating) compression is required at the output. More generally, given that the requirements are simply a multimode waveguide with irregular group-velocity spacings (thus, any multimode waveguide except for the special case of GRIN fibers, where group velocities of all modes by design are nearly identical), we expect that this work can be extended to several other wavelengths (e.g. visible or mid-IR) as well as platforms (e.g. on-chip waveguides) as a generalized scheme for nonlinear ultrafast frequency conversion in a power scalable fashion.

**Materials and Methods**

Details for replicating the experiments are provided in the supplementary material included below. It contains additional details on the experimental setup, along with discussion on the mode excitation procedure. Also the measured refractive index profiles of the higher order mode fibers are presented, together with various simulated parameters for these fibers. Finally, simulations and discussions regarding mode re-conversion to a Gaussian-like beam are provided.

**Acknowledgments**

The authors would like to thank insightful discussions with G. P. Agrawal and A. Antikainen. The authors acknowledge KMLabs for use of their ultrafast ytterbium doped fiber pump laser, and cooperation in the attendant optimization of the laser for the experiments.

**Funding:** National Institute of Health (1R21EY026410-01), AFOSR-BRI (FA9550-14-1-0165), ONR (N00014-17-1-2519), Danish Council for Independent Research and FP7 Marie Curie Actions – COFUND (DFF – 1337-00150).

# Supplementary material for:
# Discovery of Soliton Self-Mode Conversion in Multimode Optical Fibers

**S1. Materials and Methods**

A schematic representation of the experimental setup used in all of our experiments is shown in Fig. 1B. For the experimental data presented in Fig. 1, 3, 4, and 5 the pump laser was a Calmar Cazadero that emits 370 fs pulses at 1030 nm, for the data presented in Fig. 2, the pump laser was a Y-Fi laser from KMlabs, which emits 100 fs pulses at 1045 nm. The desired spatial modes of the fiber are excited with a spatial light modulator (SLM, Hamamtsu-X10468-07) that encodes an optimal spatial phase on the linearly polarized incident Gaussian-shaped beam from the laser before the light is coupled into the higher order mode (HOM) fiber. Further details on mode excitation process are described below and illustrated in Fig. S2 (32). The output of the fiber is characterized spatially using cameras (Allied-NIR-300GE, ElectroPhysics-7290A, Thorlabs-DCU224C) with appropriate spectral filters, spectrally using an optical spectrum analyzer (OSA, ANDO-AQ6317B), and temporally using an autocorrelator (AC, APE-PulseCheck).

The measured refractive index profiles of the fibers are shown in Fig. S1A – the index step is 0.04 (numerical aperture, NA=0.34), and the core diameters are 86 and 97 µm, respectively, for the two fibers used in our experiments. The inset in this plot shows a reflection microscope image of the endfacet of the fiber. The fiber shown in the red trace (HOM2) was used for results related to Fig. 2, whereas the fiber shown in the blue trace (HOM1) was used for the remaining results – note that both fibers are identical except for different core sizes. The effective index splitting between the symmetric ($LP_{0,m}$) and anti-symmetric ($LP_{1,m}$) modes in these fibers is shown in Fig. S1B. This parameter, being a proxy for linear mode stability (28), typically needs to be $\geq 5 \times 10^{-4}$, and this partly governed the choice of modes with which our experiments were conducted.

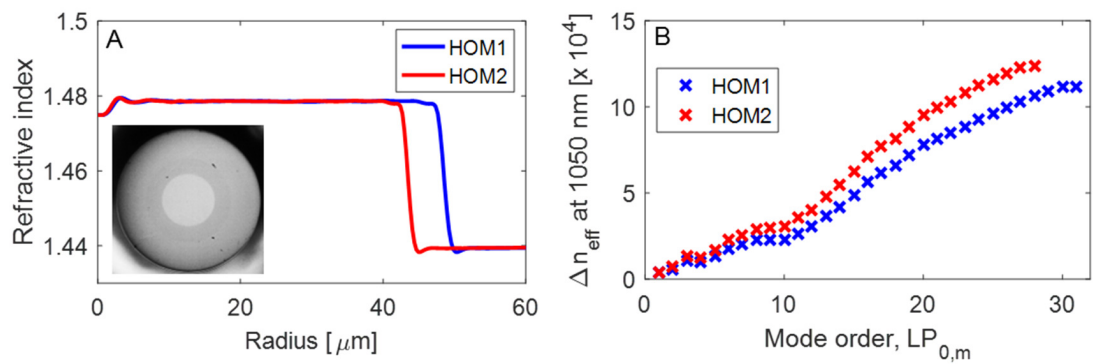

**Fig. S1 Fiber details.** (**A**) Refractive index profiles of the two HOM fibers, the inset is an image of the fiber endfacet. (**B**) Refractive index splitting ($LP_{0,m} - LP_{1,m}$) for the two fibers.

The primary aim of our mode excitation apparatus is to obtain high purity desired input modes. To achieve this, the SLM shown in Fig. 1B encodes a binary phase pattern on the incoming Gaussian beam, where the radial $\pi$ phase flips corresponds to the intensity nulls of the desired HOM, and then this SLM surface is de-magnified and imaged on to the fiber facet (*32*). Figure S2 shows one example – the green trace is the simulated field profile of the desired $LP_{0,21}$ mode in the HOM1 fiber, and also plotted (blue trace) is the field profile of an optimal Gaussian beam with the binary phase pattern imprinted on it by the SLM. The Gaussian beam waist size is chosen such that, after demagnification, its size is 40 µm at the input fiber facet. The actual size of the Gaussian beam on the SLM may thus be varied, as long as the demagnification ensures it is 40 µm in size at the fiber facet. The choice of beam waist impacts the coupling efficiency and purity of the excited mode. For excitation of the $LP_{0,21}$ mode in the HOM1 fiber using a optimally sized Gaussian beam a coupling efficiency of 72.4% was predicted and the sum of power in all others modes was -11.4 dB below that in the target mode. For excitation of the $LP_{0,19}$ mode in the HOM2 fiber the predicted coupling efficiency was 70.4% and the sum of parasitic mode excitation was -10.2 dB below that in the target mode. Similar values are obtainable for other desired modes by simply changing the phase pattern on the SLM.

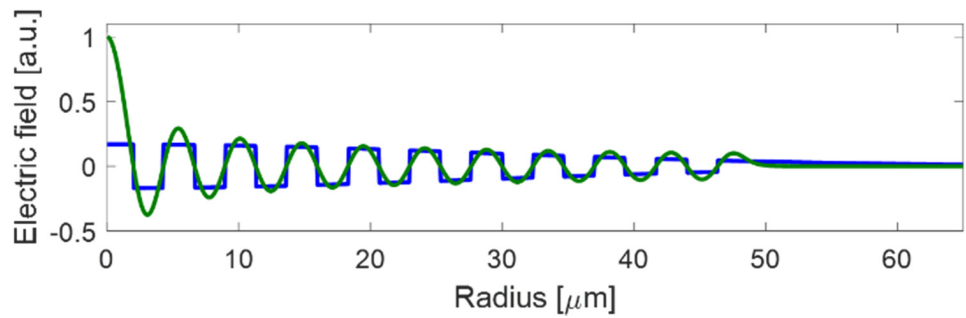

**Fig. S2 Mode excitation.** Simulated results for HOM coupling, here shown for the $LP_{0,21}$ mode into the HOM1 fiber. Green trace is the targeted $LP_{0,21}$ mode, calculated using a scalar modesolver, along with a Gaussian beam (beam waist of 40 µm at fiber facet) encoded by the SLM using a binary phase plate (concentric rings of phase 0 or $\pi$).

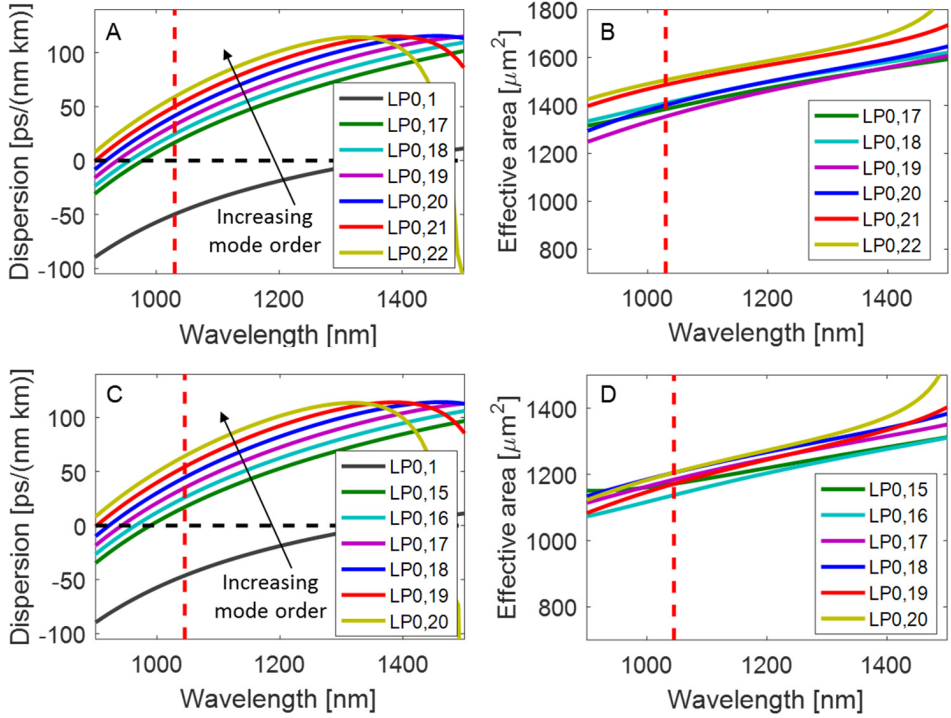

**Fig. S3 Dispersion and effective area.** (A) Dispersion and (B) effective area as a function of wavelength for a selection of modes in the HOM1 fiber shown as the blue trace in Fig. S1(a). The red dashed line indicates the pump wavelength at 1030 nm. (C) Dispersion and (D) effective area in the HOM2 fiber shown as the red trace in Fig. S1A. The red dashed line marks the pump wavelength at 1045 nm.

Figure S3 shows simulated GVD and effective area ($A_{eff}$) data for both fibers (HOM1 and HOM2) used in our experiments. The GVD is defined by

$$D = -\frac{2\pi c \beta_2}{\lambda^2}, \quad (S1)$$

where $\beta_2$ is the spectral curvature of the propagation constant (also called the dispersion parameter in some ultrafast optics literature), $c$ is the speed of light, and $\lambda$ is the wavelength. The effective area $A_{eff}$ is defined by

$$A_{eff} = \frac{\left(\int |F(x,y)|^2 dA\right)^2}{\int |F(x,y)|^4 dA}, \quad (S2)$$

where $F(x,y)$ is the transverse field distribution of the mode and $dA$ denotes integration with respect to the cross-sectional dimensions of the fiber. These parameters are important for analyzing ultrafast nonlinear processes since anomalous GVD (D>0) is a necessary condition for the existence of a soliton, and the soliton energy is given by

$$E_{soliton} = \frac{D \lambda^3 A_{eff}}{4\pi^2 c n_2 T_0}, \quad (S3)$$

where $n_2$ is the nonlinear coefficient and $T_0$ is the pulse duration. Hence, $A_{eff}$ is a proxy for the energy scalability of a fiber nonlinear process. For the experiments described in Figs. 1, 3, 4 and 5, the HOM1 fiber was used and initial pump $LP_{0,21}$ mode had $A_{eff}$ ~1486 µm² and GVD ~ 50 ps/nm km at the pump wavelength of 1030 nm. For the experiments described in Fig. 2, the HOM2 fiber was used, and the launched $LP_{0,19}$ mode had $A_{eff}$ ~ 1365 µm² and GVD ~ 37.5 ps/nm km at the pump wavelength of 1045 nm. Note, from the plots in Fig. S3 that the GVD for all subsequent modes generated by SSMC remain anomalous, and hence were able to support solitons. In addition, all their $A_{eff}$ exceeded 1100 µm². For reference, photonic crystal fibers used for dispersion tailoring in the past have $A_{eff}$ ~ 10 µm², and standard SMF in the 1000-nm region has $A_{eff}$ ~ 50 µm². Thus all participating modes had $A_{eff}$ at least one order of magnitude larger than conventional fibers used for nonlinear optics in the 1000-nm range, thereby enabling the record power levels achieved in the experiments.

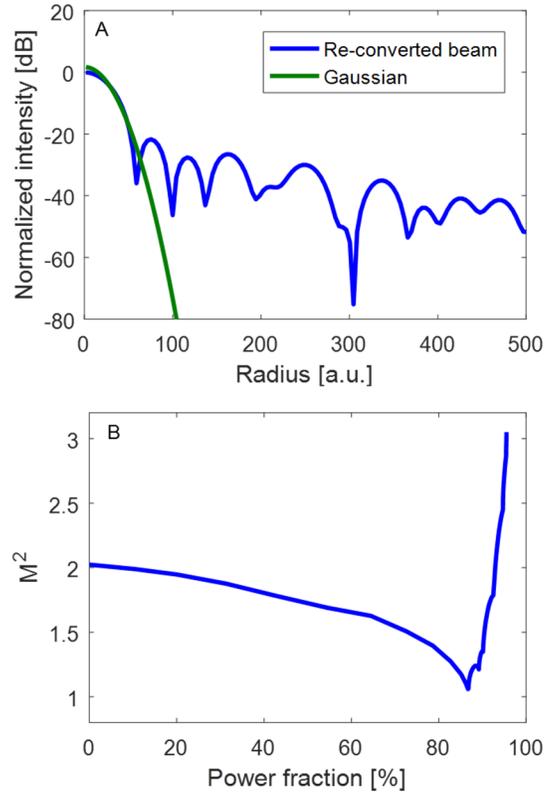

**Fig. S4 Mode re-conversion simulations.** Simulation based on Fourier beam propagation. (**A**) The axicon re-converted Gaussian-like beam, along with best fitting Gaussian beam. (**B**) The simulated $M^2$ as a function loss introduced by aperturing the beam.

The use of an axicon to convert a Gaussian beam into a Bessel-Gauss beam and vice-versa, is well known. Using Fourier beam propagation the setup shown in Fig. 5A was simulated using an axicon with a base angle of 1 degree and a 200 mm focal length lens, the starting field is an ideal $LP_{0,20}$ mode from the HOM1 fiber. Figure S4A shows the simulated re-converted Gaussian-like intensity profile obtained at the iris plane of Fig. 5A, along with an ideal Gaussian beam. Note that this plot mirrors the experimental plot of Fig. 5D, but looks substantially different since it is plotted on a logarithmic scale. As is evident, spatial aperturing is required to make the reconverted beam resemble an ideal Gaussian beam. The simulated beam quality plot of Fig. S4B (plot of $M^2$ versus the aperture size, shown as equivalent power fraction within the aperture) shows that the best theoretical outcome corresponds to a truncated re-converted beam with an $M^2 = 1.06$ when 86% of the power is transmitted.

## S2. Supplementary Figures

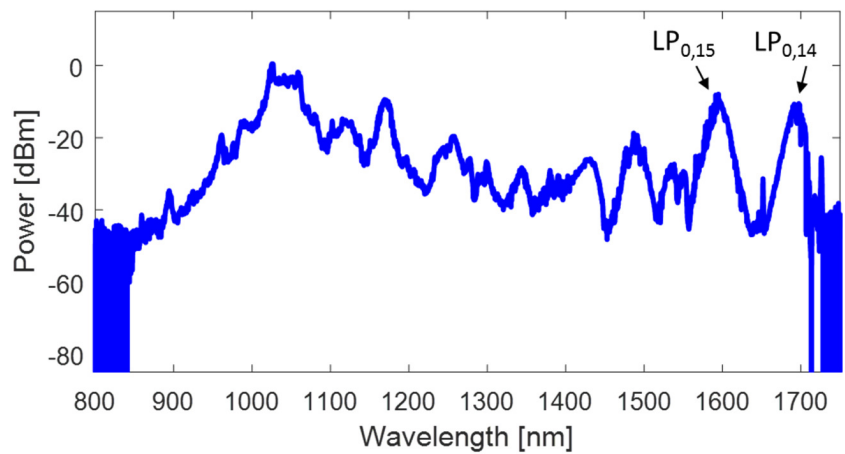

**Fig. S5 Output Spectrum.** Output spectrum after 12 m of HOM2 fiber using pump pulse energy of 100 nJ.

Figure S5 shows the output spectrum at highest pump power we employed on the full 12 m length of the fiber whose cutback results were illustrated in Fig. 2 (note that the results in Fig. 2 were obtained at lower pump power). An additional SSMC process initiates between a pulse in the $LP_{0,15}$ mode at 1595 nm and a pulse in $LP_{0,14}$ mode at 1696 nm. We expect that, with additional pump power or fiber length, full conversion would have been achieved, just as it was in all the previous cascades illustrated in Fig. 2. Moreover, nothing in our experiments suggests that SSMC would stop after this step, and hence it is conceivable that, for the right fiber length and power, even longer spectral shifts would have been feasible.